# QED Second Order Corrections on the Speed of Light at Low Temperature

H. Razmi [(1)] and M. Zamani [(2)]

Department of Physics, the University of Qom, Qom, I. R. Iran.

(1) razmi@qom.ac.ir & razmiha@hotmail.com (2) monirezamani@ymail.com

## Abstract

We want to study thermal corrections on the speed of light at low temperature considering temperature dependence of photon vacuum polarization tensor at two-loop level in the standard QED. It is found that the heat bath behaves as a dispersive medium to the propagation of light and reduces its speed proportional to the second order of temperature. Similarities and differences, with already known calculations which are based on Euler-Heisenberg Lagrangian and/or those using temperature dependent electromagnetic properties of the medium are discussed.

## 1. Introduction

The speed of light, as a fundamental constant of physics and as a speed independent of the motion of its source has a deep connection with the physical vacuum properties. The vacuum in quantum field theory is an ocean of particles that in an invisible time created and then destroyed so that they are not detectable (virtual photons); it is a stormy sea of quantum fluctuations as the physical basis for the calculations of renormalization effects and radiation corrections. There are a number of theoretically well-known and experimentally confirmed phenomena as the spontaneous emission, and the Casimir effect which are all originated from the quantum vacuum[1]. Renormalization process and radiative corrections are also due to quantum vacuum where, in QED, affect the physical (renormalized) mass of electron and its charge [2,3]. Considering the finite temperature quantum field theory, the propagators are affected by the presence of the background heat bath. The corresponding contributions are calculated either in Euclidean or Minkowski space using imaginary or real time formalism respectively [4]. In this paper, we use the real time formulation wherein energy is a continuous variables in conventional field theory which explicitly separates the $T=0$ and $T \neq 0$ components and is considerably simple [5]. The real time propagators at low temperatures (relative to the electron mass; $T \ll m_e$) can be written as [6]:

$$D_\beta^{\rho\sigma}(l) = g^{\rho\sigma}[\frac{1}{l^2+i\varepsilon} - 2\pi i \delta(l^2) n_\beta(l)], \qquad (1)$$

for photon propagator, where

$$n_\beta(l) = \frac{1}{e^{\beta|\vec{l}|} - 1} \qquad (2)$$

is the Bose-Einstein distribution and

$$S_\beta(k) = \frac{(\not{k}+m)}{k^2 - m^2 + i\varepsilon} \qquad (3)$$

for fermions (here, the Fermi-Dirac statistical factor $n_F(E) = \frac{1}{(e^{\beta E}+1)}$ in the fermion propagator at finite temperature is ignorable [7]). Thus, at $T \ll m_e$, the hot fermions contribution in background is suppressed and only the hot photons contribute. In recent decades, several papers have calculated the thermal effects on the speed of light using effective (Euler-Heisenberg) Lagrangian [8] and/or standard QED radiative corrections on the electromagnetic properties of the medium. Unfortunately, both categories deal with some difficulties to them we shall point out in **Discussion**. Here, technically working in parallel and similar to [9], considering the possibility of photon-photon scattering and examining the temperature dependence of the vacuum polarization tensor based on standard QED radiative corrections we obtain the corresponding thermal correction on the speed of light. It will be seen that "the finite temperature" (the heat bath) behaves as a dispersive medium to the propagation of light and reduces its speed proportional to the square form of temperature.

**2. Radiative correction on the vacuum polarization tensor and the "photon mass"**

The vacuum polarization tensor $\Pi_{\mu\nu}(p)$, as a second rank Lorentz tensor, can be constructed in terms of $g_{\mu\nu}$, $p_\mu p_\nu$, and scalar function $p^2$ as in the following [10]:

$$\Pi_{\mu\nu}(p) = D g_{\mu\nu} + g_{\mu\nu} p^2 \Pi^{(1)}(p^2) + p_\mu p_\nu \Pi^{(2)}(p^2) \qquad (4)$$

In the limit $p^2 \to 0$ and up to terms of higher order in $\alpha$, this yields

$$\frac{-ig_{\alpha\beta}}{p^2+i\varepsilon} \to \frac{-ig_{\alpha\beta}}{p^2+i\varepsilon} + \frac{-ig_{\alpha\mu}}{p^2+i\varepsilon}iDg^{\mu\nu}\frac{-ig_{\nu\beta}}{p^2+i\varepsilon} \qquad (5)$$

$$+ \frac{-ig_{\alpha\mu}}{p^2+i\varepsilon}iDg^{\mu\nu}\frac{-ig_{\nu\sigma}}{p^2+i\varepsilon}iDg^{\sigma\tau}\frac{-ig_{\tau\beta}}{p^2+i\varepsilon}+\ldots$$

$$\cong \frac{-ig_{\mu\nu}}{p^2-D+i\varepsilon}$$

This is just the propagator of a boson with mass $\sqrt{D}$ (a "heavy photon"). Thus, there is the following relation between the effective (dynamical) photon mass and the vacuum polarization tensor:

$$g^{\mu\nu}\Pi_{\mu\nu}(p,T) = 4D. \qquad (6)$$

**2.1. Two loop vacuum polarization**

At one loop level, because of the absence of self-interaction of photons at the approximation of low temperature limit we are considering here, the vacuum polarization tensor (in $\alpha$ order) is zero.

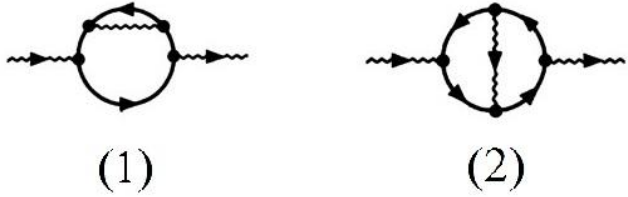

Figure 1: two loop vacuum polarization

At higher-loop level, the loop integrals involve a combination of temperature independent (cold) and temperature dependent (hot) terms which appear due to the overlapping propagator terms [11]. In order $\alpha^2$, this contribution basically comes from self mass (in Fig.1. (1) with counter

term 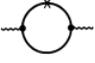) and vertex type (in Fig. 1.(2) with counter term 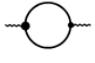 or

) of electron loop corrections inside the vacuumpolarization tensor [3]. The expression for two loop photon self energy in Fig. 1.(1) is

$$\Pi^{(1)}_{\mu\nu}(p) = e^4 \int \frac{d^4k}{(2\pi)^4} \int \frac{d^4l}{(2\pi)^4} Tr\{\gamma_\mu S_\beta(k)\gamma_\rho D_\beta^{\rho\sigma}(l) S_\beta(k+l) \qquad (7)$$
$$\times \gamma_\sigma S_\beta(k)\gamma_\nu S_\beta(p-k)\},$$

where

$$\chi^{(1)}_{\mu\nu} = Tr[\gamma_\mu(\not{k}+m)\gamma_\rho(\not{k}+\not{l}+m)\gamma^\rho(\not{k}+m)\gamma_\nu(\not{p}+\not{k}+m)] \qquad (8)$$
$$= 8[2g_{\mu\nu}m^2(m^2-k.l)$$
$$+ 3m^2\{k_\mu(p-k)_\nu + k_\nu(p-k)_\mu\}$$
$$- g_{\mu\nu}k.(p-k)\{3m^2 - 2k^4 - 4k.l\}$$
$$- m^2\{l_\mu(p-k)_\nu + l_\nu(p-k)_\mu\} + g_{\mu\nu}l.(p-k)\{m^2 - 2k^2\}]$$

while that vertex type correction to two loop photon self energy in Fig.1.(2) can be written as

$$\Pi^{(2)}_{\mu\nu}(p) = e^4 \int \frac{d^4k}{(2\pi)^4} \int \frac{d^4l}{(2\pi)^4} Tr\{\gamma_\mu S_\beta(k)\gamma_\rho D_\beta^{\rho\sigma}(l) S_\beta(k+l) \qquad (9)$$
$$\times \gamma_\nu S_\beta(k+l-p)\gamma_\sigma S_\beta(k-p)\},$$

where

$$\chi^{(2)}_{\mu\nu} = Tr[\gamma_\mu(\not{k}+m)\gamma_\rho(\not{k}+\not{l}+m)\gamma_\nu(\not{k}+\not{l}-\not{p}+m)\gamma^\rho(\not{k}-\not{p}+m)] \qquad (10)$$
$$= 8[m^2 g_{\mu\nu}(2k^2 + l^2 + p^2 - m^2 - 2k.l - l.p - 2k.p)$$
$$+ 2m^2 l_\mu(p-k-l)_\nu + 2k.(k-p)\{2(k+l)_\mu(k+l)_\nu$$
$$- (k+l)_\mu p_\nu - (k+l)_\nu p_\mu - g_{\mu\nu}[k^2 + 2k.l - k.p + l^2 - l.p]\}]$$

Due to manifest covariance in the theory, physically measurable couplings can be evaluated through contraction of vacuum polarization tensor $\Pi_{\mu\nu}$ with the metric in Minkowski space $g^{\mu\nu}$ [11]. The cold loops can then be integrated using the standard techniques of Feynman parameterization and dimensional regularization as discussed in the standard textbooks, whereas, in the same term, the evaluation of the hot loop after the cold one gives (see **Appendix**)

$$g^{\mu\nu}\Pi_{\mu\nu}^{(1)+(2)}(p,T) = \frac{\alpha^2 T^2}{3} \qquad (11)$$

About this point that why the rhs in (11) is independent of $p$, we should explain that there are a number of terms depending on $p$ but all are coefficients of (negligible) higher orders of $T$. Using (6)

$$D = \frac{\alpha^2 T^2}{12} \qquad (12)$$

Thus, the photon "mass" is

$$m_{"photon"} = \sqrt{D} = \frac{\alpha T}{2\sqrt{3}} \qquad (13)$$

**3. Radiative corrections on the speed of light**

To keep gauge invariance of QED, instead of considering a "mass" for photon, we prefer to correct the speed of light; this is because it is in a "dispersive" medium (heat bath) which reduces its speed from 1 ($c = 1$) to $v$. Thus, using (13), the correction on photon energy is found as:

$$\Delta E = \frac{\frac{\alpha T}{2\sqrt{3}}}{\sqrt{1-v^2}} \qquad (14).$$

Indeed, we have considered this qualitative feature that the heat bath behaves like a dispersive medium for photon and thus, because of screening effect due to vacuum polarization (virtual pairs of electron-positron), reduces its speed. The energy correction due to scattering from the virtual electron-positron pairs should be of the order of $m_e$:

$$\Delta E \sim 2m_e \qquad (15).$$

Comparing (14) and (15):

$$\frac{\frac{\alpha T}{2\sqrt{3}}}{\sqrt{1-v^2}} \sim 2m_e \qquad (16)$$

or

$$v^2 \sim 1 - \frac{\alpha^2 T^2}{48 m_e^2} \qquad (17).$$

So, considering the approximation regime $T \ll m_e$, it is found that:

$$v \sim 1 - \frac{\alpha^2 T^2}{96 m_e^2} \qquad (18).$$

**4. Discussion**

In some papers, thermal effects on the speed of light have been already considered based on Euler-Heisenberg Lagrangian [8]:

$$\mathcal{L}_{eff} = -\frac{1}{4} F^{\mu\nu} F_{\mu\nu} - \frac{\alpha^2}{90 m^4} \left[ \left( F^{\mu\nu} F_{\mu\nu} \right)^2 + \frac{7}{4} \left( F_{\mu\nu} \tilde{F}^{\mu\nu} \right)^2 \right] \qquad (19);$$

the final results are a reduction in the speed of light proportional to the fourth order of temperature [12]

$$c = 1 - \frac{44\pi^2 \alpha^2}{2025} \frac{T^4}{m^4} \qquad (20).$$

One serious criticism on these calculations is that, applying the Euler-Lagrange equation to the starting Lagrangian, one cannot reach the standard wave equation but deals with a nonlinear equation by which we cannot introduce the "velocity" (speed) in its standard form:

$$-\frac{1}{c^2} \frac{\partial^2 \vec{A}}{\partial t^2} \frac{\left[1 + \frac{\alpha^2}{10m^4} \frac{1}{c^2} (\frac{\partial \vec{A}}{\partial t})^2\right]}{\left[1 + \frac{\alpha^2}{30m^4} \frac{1}{c^2} (\frac{\partial \vec{A}}{\partial t})^2\right]} + \vec{\nabla}^2 \vec{A} = 0 \qquad (21).$$

Some other papers [6, 9, 11, 13] work with the standard QED radiative corrections and try to find the electron charge renormalization, the electric permittivity, and the magnetic permeability in terms of temperature. Although there is no direct calculation of the speed of light in these papers, it can be simply shown that the resulting speed based on the calculated renormalized electron charge (and/or electromagnetic properties of the medium (heat bath)) is greater than 1 (c=1)! Considering no screening of magnetic fields due to the transverse nature of $\Pi_{ij}$ implied by gauge invariance [14] and the following result from [9]:

$$Z_3 = 1 + \frac{\alpha^2 T^2}{6m^2} \qquad (22),$$

it is found

$$v = \frac{1}{\sqrt{\varepsilon}} = \sqrt{Z_3} = 1 + \frac{\alpha^2 T^2}{12m^2} \qquad (23).$$

Indeed, it seems the correct way of having a reasonable method and result is to consider the presence of the thermal heat bath as a dispersive medium and then have had a kinematic calculation of the reduced speed of light due to its dynamically generated mass. Of course, as we have already mentioned, we don't consider the "mass" of photon as its rest mass (or any other similar quantity) to prevent the possibility of destroying gauge invariance of the theory.

**References**

[1] P. W. Milonni, *The Quantum Vacuum: an introduction to quantum electrodynamics* (Academic Press, 1993).

[2] F. Mandl and G. Shaw, *Quantum Field Theory*, (Wiley, 2010) chapters 9 & 10.

[3] C. Itzykson and J. Zuber, *Quantum Field Theory* (McGraw-Hill, 1987).

[4] T. Matsubara, *Progress of Theoretical Physics* **14** (1955) 351; A. Das, *Finite Temperature Field Theory* (World Scientific, 1997).

[5] J. F. Donoghue, B. R. Holstein and R. W. Robinett, *Ann. Phys.* **164** (1985) 233-276; J. F. Donoghue and B. R. Holstein, *Phys. Rev.*D **28** (1983) 340-348.

[6] K. Ahmed and Samina S. Masood, *Ann. Phys.* **207** (1991) 460-473.

[7] K. Ahmed and Samina Saleem, *Phys. Rev.D* **35** (1987) 1861-1871.

[8] W. Heisenberg and H. Euler, *Z. Phys.* **98** (1936) 714; H. Euler, *Ann. Phys.* **26** (1936) 398; W. Greiner and J. Reinhart, *Quantum Electrodynamics* (Springer, New York, 2003) chapter 7.

[9] Samina S. Masood and Mahnaz Q. Haseeb, *Int. J. Mod. Phys. A* **23** (2008) 4709-4719.

[10] W. Greiner and J. Reinhart, *Quantum Electrodynamics* (Springer, New York, 2003) chapter 5.

[11] Mahnaz Haseeb, Samina S. Masood, hep-th/1110.3447.

[12] X. Kong and F. Ravandal, *Nucl. Phys. B* **526** (1998) 627-656; F. Ravandal, hep-ph/9709220, (Contributed talk at "Strong and Electroweak Matter 97", Eger, Hungary, May 21-25, 1997).

[13] A. Weldon, *Phys. Rev. D* **26** (1982) 1394-1407; R. Tarrach, Phys. Let. B**133** (1983) 259-261.

[14] R. M. Woloshyn, *Phys. Rev. D* **27** (1983) 1393-1395.

# Appendix

At low temperature, the hot contribution comes from the photon background only. Therefore, for example, Eq. (7) for Fig.1.(1) simplifies to

$$\Pi^{(1)}_{\mu\nu} = \Pi^{(1)}_{\mu\nu}(p, T=0)$$
$$-\frac{2\pi i e^4}{(2\pi)^8} \int d^4k \int d^4l \frac{\chi^{(1)}_{\mu\nu} \delta(l^2) n_\beta(l)}{[k^2 - m^2 + i\varepsilon]^2 [(k+l)^2 - m^2 + i\varepsilon][(p-k)^2 - m^2 + i\varepsilon]} \quad (A.1),$$

then

$$g^{\mu\nu} \Pi^{(1)}_{\mu\nu}(p, T) = -2\pi i e^4 \int \frac{d^4k}{(2\pi)^4} \int \frac{d^4l}{(2\pi)^4} \times$$
$$\frac{g^{\mu\nu} \chi^{(1)}_{\mu\nu} \delta(l^2) n_\beta(l)}{[k^2 - m^2 + i\varepsilon]^2 [(k+l)^2 - m^2 + i\varepsilon][(p-k)^2 - m^2 + i\varepsilon]} \quad (A.2)$$

where

$$g^{\mu\nu} \chi^{(1)}_{\mu\nu} = 8[(2m^4 - m^2 l.p) + 3m^2(p_\mu + l_\mu)k^\mu + (-3m^2 + 2l.p)k^2 + 2k^2(k.p - k.l) - 2k^4] \quad (A.3).$$

Integrating over the cold loop using Feynman parameterization and dimensional regularization, one gets

$$g^{\mu\nu} \Pi^{(1)}_{\mu\nu}(p, T) = \frac{e^4}{16\pi^5} \int_0^1 dx \int_0^x dy \int_0^y dz \int d^4l \, \delta(l^2) n_\beta(l)$$
$$\left[ \frac{1}{[m^2 + 2l.pz(y-z)]^2} \left\{ 2m^4 - m^2(l.p)\left[1 + 3y - 6z + 6z^2 - 6yz\right] \right. \right.$$
$$\left. - 2(l.p)^2 \left[yz - z^2 + 4z^4 + 4z^2y^2 - 8z^3y\right] \right\}$$
$$\left. - \frac{1}{[m^2 + 2l.pz(y-z)]} \left\{ -6m^2 + (l.p)\left[1 - 12z^2 + 12yz\right] \right\} - \frac{12}{\varepsilon} \right] \quad (A.4)$$

and

$$g^{\mu\nu}\Pi^{(1)}_{\mu\nu}(p,T) = \frac{\alpha^2}{\pi^2}\int_0^1 dx \int_0^x dy \int_0^y dz \int_{-1}^1 d(\cos\theta) \int d|\vec{l}|\, \frac{|\vec{l}|}{e^{\beta|\vec{l}|}-1}$$

$$\times \left\{ 2m^4\left[\frac{1}{a_+^2}+\frac{1}{a_-^2}\right]\right.$$

$$-m^2|\vec{l}|[1+3y-6z+6z^2-6yz]\left[\frac{p_0-|\vec{p}|\cos\theta}{a_+^2}-\frac{p_0+|\vec{p}|\cos\theta}{a_-^2}\right]$$

$$-2|\vec{l}|^2[yz-z^2+4z^4+4z^2y^2-8z^3y]\left[\frac{(p_0-|\vec{p}|\cos\theta)^2}{a_+^2}+\frac{(p_0+|\vec{p}|\cos\theta)^2}{a_-^2}\right]$$

$$+6m^2\left[\frac{1}{a_+}+\frac{1}{a_-}\right]$$

$$\left. -|\vec{l}|[1-12z^2+12yz]\left[\frac{p_0-|\vec{p}|\cos\theta}{a_+}-\frac{p_0+|\vec{p}|\cos\theta}{a_-}\right]-\frac{12}{\varepsilon}\right\} \qquad (A.5)$$

where $a_\pm = m^2 \pm 2z(y-z)|\vec{l}|(p_0 \mp |\vec{p}|\cos\theta)$.

It should be noted that the $\frac{1}{\varepsilon}$ contribution is cancelled by the counter term

diagram 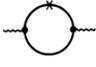. After a lengthy calculation of the integrals and regardless of the higher powers of $T$, one finds

$$g^{\mu\nu}\Pi^{(1)}_{\mu\nu}(p,T) = \frac{8\alpha^2 T^2}{9} \qquad (A.6).$$

Similarly

$$g^{\mu\nu}\Pi^{(2)}_{\mu\nu}(p,T) = -\frac{5\alpha^2 T^2}{9} \qquad (A.7).$$

Then

$$g^{\mu\nu}\Pi^{(1)+(2)}_{\mu\nu}(p,T) = \frac{\alpha^2 T^2}{3} \qquad (A.8).$$